\documentclass[%
 reprint,
 amsmath,amssymb,
 aps,
pra
]{revtex4-2}

\usepackage{graphicx}
\usepackage{dcolumn}
\usepackage{bm}
\usepackage{hyperref}
\usepackage{amsmath}

\let\oldphi=\phi
\renewcommand{\phi}{{\oldphi^{p,q}_\hbar}}
\newcommand{\Q}{Q_\psi^\hbar}
\newcommand{\R}{\mathbb{R}}

\usepackage{derivative}

\begin{document}

\title{Forward-Time Equivalent of a ``Retrocausal" Diffusion\\ Hidden Variable Model for Quantum Mechanics}

\author{William S. DeWitt}
\email{wsdewitt@uw.edu}
 \affiliation{Department of Genome Sciences, University of Washington.}
\author{Benjamin H. Feintzeig}%
 \email{bfeintze@uw.edu}
\affiliation{Department of Philosophy, University of Washington
}%

\begin{abstract}
A recently proposed stochastic hidden variable model for quantum mechanics has been claimed to involve ``retrocausality" due to the appearance of equations of motion with future-time boundary conditions.
We formulate an equivalent system of forward-time equations of motion that gives rise to the same trajectories as solutions, but involves only initial-time boundary conditions.
The forward-time dynamics involves a guidance term for the dynamical variables, determined by the phase-space distribution corresponding to a quantum wavefunction.
We show, however, that this particular guidance term can be recovered as the mean-field limit of averaged pairwise interactions among an ensemble of finitely many particles.   The results establish a general trade-off in stochastic models for quantum mechanics between reverse-time and guiding wave formulations.
\end{abstract}

\maketitle

\section{Introduction}

In a series of recent papers, Drummond and Reid \citep{DrRe20,Dr21,DrRe21} have proposed a new hidden variable model for quantum mechanics and argued that it helps solve interpretive and foundational problems associated with quantum measurement \citep{JoThHaFuTeDrRe24,McTeDrRe24}.
Drummond and Reid describe their model as including ``retrocausal" effects because the defining stochastic differential equations (SDEs) involve a boundary condition at a future time, resulting in a reverse-time diffusion process.
In this letter, we show that ``retrocausality" need not be understood as an inherent feature of the model.  Using a classic result on time-reversed diffusion processes \citep{An82} (of recent acclaim for its key role in score-based generative modeling through SDEs \cite{So20}), we formulate an equivalent forward-time stochastic process that gives rise to the same solutions as the model from Drummond and Reid.
Our dynamical model has some resemblance to older pilot wave hidden variable theories \citep{Bo52,Ne66}, crucially adjusted to provide a stochastic dynamics on phase space.
Our stochastic formulation on phase space generalizes pilot wave theories on phase space \citep{dP96} and also admits an interpretation in terms of \emph{propagation of chaos} in an interacting particle system \cite{ChDi22a,ChDi22b,Sz91} that resembles many-interacting-worlds proposals \citep{HaDeWi14}.
Our forward-time equations, as well as the approximation by finite-particle systems that we propose, also give rise to new methods of computing the allowed trajectories in Drummond and Reid's model, even in cases where directly solving their reverse-time equations may prove difficult.

Our goal here is not to adjudicate between the ``retrocausal" model of Drummond and Reid and our corresponding forward-time guiding wave formulation.  We believe that both interpretations of quantum mechanics are interesting and worthy of investigation.  Instead, our aim is to clarify the connections between different formulations and interpretations of quantum theory, as well as the existing precedents for what seem to be recent new developments.  One aspect of our finding is that there are deep connections between the mathematical literature concerning time reversal of stochastic processes and applications to the physics of hidden variable models for quantum mechanics.  We hope to show that these rich connections, as yet underexplored, provide fertile ground for further investigation in the foundations of quantum mechanics.

\section{``Retrocausal'' Diffusion Model}

We summarize Drummond and Reid's ``retrocausal" stochastic hidden variable model through the concrete example of a single particle quantum system with state space consisting of square-integrable normalized wavefunctions in one dimension.  For further detail, see also the discussion of \citet{Fr24}.
The model employs the \emph{Husimi Q-function} $\Q$, an integrable function on the two-dimensional phase space with canonical coordinates $(p,q)$.
The Husimi function $\Q$ is defined for any wavefunction $\psi$ by
\begin{equation}
    \Q(p,q) = \frac{1}{2\pi\hbar}\left|\left\langle\phi|\psi\right\rangle\right|^2,
\end{equation}
where
\begin{equation}
    \phi(x) = \frac{1}{(\pi\hbar)^{1/4}}e^{-\frac{ipq}{2\hbar}}e^{\frac{ipx}{\hbar}}e^{-\frac{(x-q)^2}{2\hbar}}
\end{equation}
is a \emph{coherent state}.
The Husimi function $\Q$ is a probability density on the phase space corresponding with the Berezin quantization scheme.
Berezin quantization \citep{La17} associates a suitable phase space function $f$ (e.g., smooth and compactly supported) with an operator or ``symbol" $\mathcal{Q}_\hbar(f)$, determined by the prescription
\begin{equation}
    \mathcal{Q}_\hbar(f) = \int_{\R^2} \frac{\odif{p,q}}{2\pi\hbar}f(p,q)\left|\phi\right\rangle\left\langle\phi\right|,
\end{equation}
where the integral is understood in the weak operator sense.
For each suitable phase space function $f$, the operator $\mathcal{Q}_\hbar(f)$ is compact and satisfies
\begin{equation}
    \left\langle \psi|\mathcal{Q}_\hbar(f)|\psi\right\rangle = \int_{\R^2} f(p,q)\,\Q(p,q)\odif{p,q}
\end{equation}
for all wavefunctions $\psi$, thus recovering quantum mechanical expectation values for the operator $\mathcal{Q}_\hbar(f)$ from the classical expectation value of $f$ relative to the Husimi probability density.
Those unfamiliar with the Husimi function can compare its role as a phase space probability density with that of the perhaps more familiar Wigner function \citep{Wi32,Mo49}, which is a phase space quasi-probability density associated with the Weyl ``symbols" determined by the Weyl quantization scheme \citep{We50,La98b}.

Drummond and Reid's model \citep{Dr21,DrRe20} is motivated by the observation that, for a class of Hamiltonians, the Schr\"odinger equation $\psi(t) = e^{-iHt/\hbar}\psi_0$ (for initial condition $\psi(0) = \psi_0$) gives rise to a generalized Fokker-Planck equation for the Husimi function (which here, and in what follows, is understood as a flow of phase-space densities due to the time dependence of $\psi$):
\begin{multline}
\label{eq:Fokker-Planck}
    \partial_t\Q = -\partial_p (a_1\Q) - \partial_q(a_2\Q)\\
    + \frac{1}{2}\left(\partial_p^2 \left(b^2\Q\right) - \partial_q^2\left(b^2\Q\right)\right).
\end{multline}
Here,  $a_1, a_2$ are positive \emph{drift coeffecients}, and $b$ is a positive \emph{diffusion coefficient} (but note the negative sign on the $\partial_q^2$ term).  All of the drift and diffusion coefficients may vary as a function of the phase-space variables.  Eq.~\eqref{eq:Fokker-Planck} follows for a class of Hamiltonians with potential up to fourth order in $x$; we do not specify the exact conditions on the Hamiltonian, however, for reasons of space.
The Fokker-Planck type Eq.\ \eqref{eq:Fokker-Planck} resembles standard diffusion dynamics, but, whereas dissipative diffusion involves positive diffusivity, the time symmetry of the quantum dynamics in the Schr\"odinger equation force the use of some nonstandard negative entries to maintain traceless diffusivity.

Drummond and Reid's model in principle allows for a traceless matrix of diffusion coefficients acting on second-order derivatives of phase-space variables, where the matrix may be diagonal in a coordinate system different from the standard one for $(p,q)$.
We restrict ourselves to the simpler form in Eq.\ \eqref{eq:Fokker-Planck} because this encompasses our examples and suffices to illustrate the main ideas.

Drummond and Reid suggest to interpret the Fokker-Planck type Eq.\ \eqref{eq:Fokker-Planck} as the bulk evolution of a probability density corresponding to some underlying microdynamics for the ``hidden variables" $p$ and $q$ represented by an SDE.
They propose the coupled SDEs written in integral form as
\begin{align}
\label{eq:pSDE}
    p(t) &= p(t_0) + \int_{t_0}^t a_1\odif{t'} + \int_{t_0}^t b \odif{w_1},\\
    \label{eq:qSDE}q(t) &= q(t_f) - \int_t^{t_f}a_2 \odif{t'} - \int_{t}^{t_f} b \odif{w_2},
\end{align}
where $w_1$ and $w_2$ are independent Gaussian noise processes characterized by
\begin{equation}
    \left\langle w_i(t)w_j(t')\right\rangle = \delta_{ij}\,\delta(t-t'), \quad i,j = 1,2
\end{equation}
Above and in what follows, we use $\langle X \rangle_\mu$ to denote the expectation value of $X$ with respect to a measure $\mu$, and we drop the subscript when the choice of measure is clear.

While we have stated these SDEs in their integral form, one can also state a corresponding differential form.  To do so, define the reverse-time stochastic process $\overline{q}(t) = q(t_f-t)$.  Then Eqs.~\eqref{eq:pSDE}-\eqref{eq:qSDE} are equivalent to the following equations for increments:
\begin{align}
\odif{p}(t) &= a_1\odif{t} + b \odif{w}_1\\
\odif{\overline{q}}(t) &= a_2\odif{t} + b\odif{w}_2.
\end{align}
The equations are, in general, coupled through the functional form of the drift terms, since $a_1$ may depend on $q$ and $a_2$ may depend on $p$.
The Langevin-type stochastic Eqs.\ \eqref{eq:pSDE}-\eqref{eq:qSDE} involve boundary conditions consisting of both an \emph{initial} time $t_0$ and a \emph{future} time $t_f$, thus leading Drummond and Reid to interpret them as representing a ``retrocausal" model of quantum theory.
According to Drummond and Reid, these SDEs recover the bulk evolution of the distribution $\Q$ in Eq.\ \eqref{eq:Fokker-Planck}, thus yielding the standard quantum mechanical predictions.

We mention briefly that there is some precedent for using the Husimi function as a basis for a probabilistic interpretation of quantum mechanics in other works \citep{Bo56,Mi84,Ap00,BeGaYo13}, which Drummond and Reid's model builds upon.

\section{Equivalent Forward-Time Process}

We motivate our alternative approach by noticing that the ``retrocausal" aspects of Drummond and Reid's model do not appear at all for certain systems.
Well-known results concerning Berezin quantization \citep[e.g.,][Cor.\ II.2.4.5]{La98b} imply immediately that for a class of Hamiltonians up to quadratic in the phase space variables $(p,q)$, the Fokker-Planck equation can be recovered from a completely deterministic microdynamics.

For example, for a free particle Hamiltonian
\begin{equation}
    H = -\frac{\hbar^2}{2m}\partial_x^2,
\end{equation}
one finds the Fokker-Planck equation for $\Q$ in the form
\begin{equation}
\label{eq:free_FP}
    \partial_t \Q = -\frac{p}{m}\partial_q\Q.
\end{equation}
This can be recovered from the microdynamics
\begin{align}
     p(t) &= p(t_0),\\
    q(t) &= q(t_0) + \int_{t_0}^t\frac{p(t')}{m}\odif{t'} \nonumber\\
    &= q(t_0) + \frac{p(t_0)}{m}(t-t_0)
\end{align}
which does not require a future boundary condition.
This example raises the question: is the use of a future boundary condition \emph{necessary} to reproduce a Fokker-Planck equation for $\Q$ as in Drummond and Reid's model?

While Drummond and Reid analyze only one possible equation for the microdynamics underlying the bulk Fokker-Planck type evolution of the probability density in Eq.\ \eqref{eq:Fokker-Planck}, generally such microdynamics may be non-unique.
In this case, we will now define an alternative microdynamics reproducing the same bulk evolution.
Whereas Drummond and Reid's microdynamics involves a reverse-time diffusion process for one of the dynamical variables, our proposed equivalent microdynamics involves only forward-time evolution equations.
To proceed, we invoke a classic result concerning reverse-time diffusion dynamics from Anderson \citep{An82}, which states that every suitable diffusion process has an equivalent time-reversed process giving rise to a one-to-one correspondence between forward-time and backward-time trajectories  (See also \citep{Fo85,FoWa86}).
By a ``suitable" diffusion process, we mean one whose Fokker-Planck equation has a well-defined initial value problem giving rise to a unique smooth evolution of the probability density, as well as a unique smooth joint evolution with the modified noise term below.

In our case, we carry over Drummond and Reid's microdynamics in Eq.\ \eqref{eq:pSDE} for the evolution of $p$, but we reverse the time in Eq.\ \eqref{eq:qSDE} for the evolution of $q$ to obtain the forward-time SDE in integral form:
\begin{equation}
\label{eq:forwardqSDE}
    q(t) = q(t_0) + \int_{t_0}^t \overline{a}_2\odif{t'} + \int_{t_0}^t b  \odif{\overline{w}_2}.
\end{equation}
In this equation, we employ the new drift term
\begin{equation}
\label{eq:guidancedrift}
    \overline{a}_2(p,q) = a_2(p,q) + \frac{\partial_q \left(b^2(p,q) \ \mu_t(q)\right)}{\mu_t(q)},
\end{equation}
where
\begin{equation}
    \mu_t(q) = \int_\R \Q(p,q) \odif{p}
\end{equation}
is the marginal probability density over the variable $q$ at time $t$.
The new diffusion term involves the modified Gaussian noise process $\overline{w}_2$ with increments that relate to those of the forward process according to
\begin{equation}
    \odif{\overline{w}_2}(t) = \odif{w_2}(t) + \frac{\partial_q \left(b(p,q) \ \mu_t(q)\right)}{\mu_t(q)}\odif{t},
\end{equation}
which likewise is characterized by
\begin{equation}
    \left\langle \overline{w}_2(t)\overline{w}_2(t')\right\rangle = \delta(t-t').
\end{equation}
Again, while we have stated Eq.~\eqref{eq:forwardqSDE} in its integral form, one can also state a corresponding differential form:
\begin{align}
    \odif{q}(t) = \overline{a}_2\odif{t} + b\odif{\overline{w}}_2.
\end{align}
Crucially, the forward-time microdynamics defined in Eq.\ \eqref{eq:forwardqSDE} is a forward-time It\^o stochastic process as long as (i) $\mu_t(q)$ satisfies suitable regularity or smoothness assumptions required for the existence and uniqueness of a smooth solution to the Fokker-Planck Eq.\ \eqref{eq:Fokker-Planck}, and (ii) there is a unique smooth solution to the Kolmogorov equation corresponding to the evolution of the joint probability density associated with $q(t)$ and $\overline{w}_2(t)$.  Under these conditions, the main theorem of \citet{An82} establishes that Eq.~\eqref{eq:qSDE} holds if and only if Eq.~\eqref{eq:forwardqSDE} holds.

Notice that the new drift term in Eq.\ \eqref{eq:guidancedrift} couples to the marginal probability distribution in a similar form to pilot wave guidance equations.
For instance, the de Broglie-Bohm pilot wave theory \citep{Bo52,DuGoZa92} contains a velocity term of the form $\Im(\partial_x\psi/\psi)$ for a pilot wavefunction $\psi(x)$, and likewise Nelson's stochastic mechanics \citep{Ne66} contains a velocity term of the form $\partial_x\rho/\rho$ for a probability density $\rho(x) = \left|\psi(x)\right|^2$.
Thus, Eq.\ \eqref{eq:forwardqSDE} can be understood as a stochastic guidance equation for a particle undergoing diffusion guided by the Husimi function.
The familiar form of these new drift terms in pilot wave theories can now be understood as arising from the general considerations regarding the generation of time-symmetric differential equations using the procedure from Anderson \citep{An82} for defining time-reversed processes.

This perspective on guidance equations suggests further questions about existing pilot wave models for quantum mechanics.
Can one characterize a reverse-time equation equivalent to the de Broglie-Bohm dynamics? Or can one characterize a reverse-time equation equivalent to Nelson's mechanics?
The results of \citet{An82} guarantee these reverse-time equations must exist.  
Comparing them to the model of \citet{DrRe20} may lead to further insight.

Should the forward-time guiding wave formulation of Eq.~\eqref{eq:guidancedrift} be \emph{preferred} over Drummond and Reid's ``retrocausal" model of Eq.~\eqref{eq:qSDE}?
\citet{Fr24} has argued that Drummond and Reid's model does not involve true reverse-time causation---if this is correct, then even a desire to avoid retrocausality does not entail a problem with Drummond and Reid's approach.
Our aim is not to argue against the reverse-time formulation, or even in favor of the forward-time formulation as an interpretation of quantum theory.
We only claim to have shown that there is a trade-off between our forward-time guiding wave equations and the reverse-time formulation of Drummond and Reid's SDEs, regardless of whether the reverse-time equations involve genuine retrocausality.
Even setting aside interpretive issues, the reverse-time formulation for general couplings of $p$ and $q$ may be very difficult to solve or simulate.
So we have at least provided a practical tool in the forward-time guiding wave formulation for further investigations of stochastic diffusion models for quantum mechanics.

\section{Mean-Field Interpretation}

One potential virtue of the forward-time stochastic guidance diffusion model in Eq.\ \eqref{eq:forwardqSDE}-\eqref{eq:guidancedrift} is that it is of McKean-Vlasov type \citep{McK66,ChDi22a}, with coefficients that depend on the distribution of the solution of the Fokker-Planck equation.
Hence, the model can be interpreted as the mean-field limit of an ensemble of $N$ particles $\left(q^N_i(t)\right)_{i=1}^N$ interacting weakly via their empirical measure
\begin{equation}
    \mu^N_t = \frac{1}{N}\sum_{i=1}^N \delta_{q^N_i(t)},
\end{equation}
and driven by independent noise processes $\left(\overline{w}^N_{2,i}(t)\right)_{i=1}^N$ of intensity $b$ characterized by
\begin{equation}
    \left\langle \overline{w}^N_{2,i}(t)\overline{w}^N_{2,j}(t')\right\rangle = \delta_{ij}\,\delta(t-t'), \quad i,j = 1,\dots,N.
\end{equation}
Since the positions $q^N_i(t)$ are random variables sampled from a distribution satisfying an SDE, the empirical measure $\mu^N_t$ is a \emph{measure-valued} stochastic process.

Explicitly, we specify the so-called \emph{$N$-system} $\left(q^N_i(t)\right)_{i=1}^N$ to satisfy $N$ coupled SDEs
\begin{equation}
\label{eq:NsystemSDE}
    q_i^N(t) = q_i^N(t_0) + \int_{t_0}^t \overline{a}^N_2\odif{t'} + \int_{t_0}^t b  \odif{\overline{w}^N_{2,i}}.
\end{equation}
The N-system SDEs are coupled via the appearance of the empirical measure in their drift term
\begin{equation}
\label{eq:interactingSDE}
\bar a^N_2(p,q) = a_2(p,q) + \frac{\partial_q\left(b^2(p,q)\,\tilde\mu_t^N(q)\right)}{\tilde\mu_t^N(q)},
\end{equation}
with \emph{mollified} (kernel smoothed) empirical measure
\begin{align}
    \tilde\mu_t^N(x) = \left(\mu_t^N \ast K^N\right)(x) = \int K^N(x-y) \mu_t^N(y) \odif{y}.
\end{align}
Here, $\ast$ denotes convolution and we choose $K^N$ as a positive symmetric \emph{mollifier} (smoothing kernel) with bandwidth $\frac{1}{N}$.  As a concrete example, $K^N(\cdot)$ could be a Gaussian with variance $\frac{1}{N^2}$ and mean $0$.

If the initial states $\left(q^N_i(0)\right)_{i=1}^N$ are sampled from an iid (or more generally, exchangeable) distribution, then it follows from standard results on propagation of chaos for McKean-Vlasov systems that, in the limit $N\to\infty$ (so long as relevant derivatives are bounded), we have $\mu^N_t\to\mu_t$ in the sense of convergence in the Wasserstein metric \citep[][Thm.\ 3.1]{ChDi22b}, which also implies weak* convergence, or convergence in distribution.
Further, a defining feature of the mollifier is that $K^N\to\delta$ (weak* convergence as tempered distributions), so we also have $\tilde\mu_t^N\to\mu_t$.
Thus we have $\bar a^N_2\to\bar a_2$ (convergence in distribution), matching the guidance term in Eq.\ \eqref{eq:guidancedrift}.

In this mean-field limit, the stochastically evolving empirical measure $\mu^N_t$ converges to the \emph{deterministically} evolving measure $\mu_t$, whose flow is determined by the Fokker-Planck Eq.\ \eqref{eq:Fokker-Planck}.
Furthermore, the $N$-system \emph{decouples}, so that each particle in the ensemble can be treated as evolving independently only taking into account interactions with the background field $\mu_t$.
These features characterize \emph{propagation of chaos} (See \citep{ChDi22a,ChDi22b,Sz91} for comprehensive reviews of the theory and many applications of this vast topic).

One might wonder whether the $N$-system we propose could provide an alternative physical interpretation, or whether it can only serve as a formal or heuristic analogy.
While of course an interpretation of the $N$ system as providing a formal or heuristic analogy is possible, we note that the picture provided by our mean-field interpretation also bears an interesting resemblance to the proposal of \citet{HaDeWi14}, which aims to recover quantum dynamics from the interaction of many classical worlds.
We suggest that one might develop a full physical interpretation of the $N$-system approximation by treating the interacting particles as something like worlds.
For the purposes of the current paper, we remain agnostic between these interpretive options.
A complete discussion would require weighing the costs and benefits of ``many-worlds" and ``single-world" interpretations.

Our discussion of the connection between propagation of chaos and reverse-time diffusion is anticipated by \citet{Na93}, who applied similar methods directly to the evolution of the wavefunction, treating the Schr\"odinger equation as a type of diffusion equation (See also \citep{NaTa87a,NaTa87b}).
Here, we have extended these methods from treating wavefunction amplitudes to directly treating probabilities represented by Husimi functions.

\section{Example: Amplifier Dynamics}
\label{sec:example}

\begin{figure*}[t]
    \centering
    \includegraphics[width=1\textwidth]{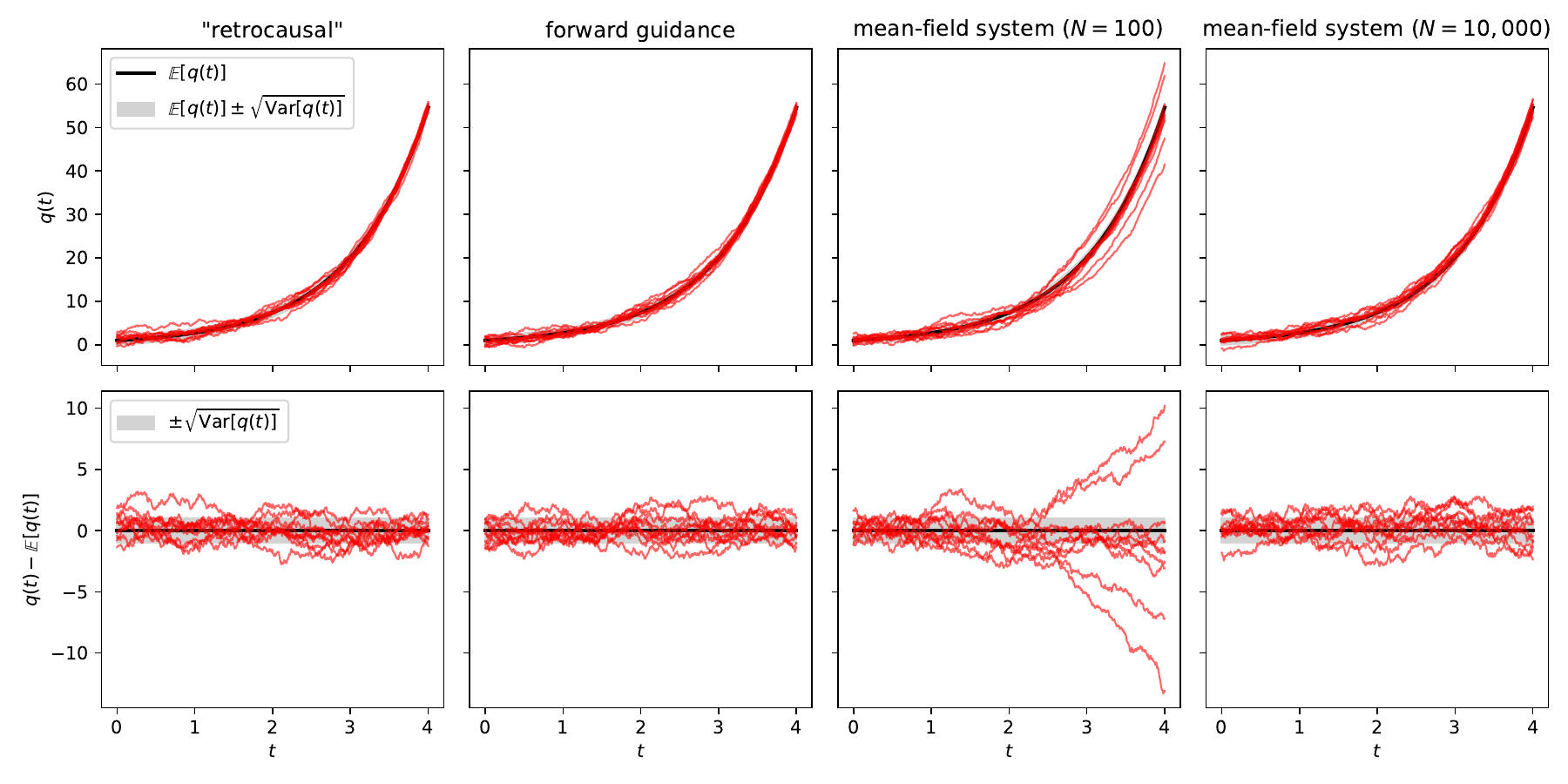}
    \vspace{-20pt}
    \caption{Comparisons of numerical solutions for the various formulations of microdynamics for the amplifier model of \S~\ref{sec:example}.  From left to right, we include the \emph{``retrocausal"} model from Drummond and Reid \citep{DrRe20} in Eq.~\eqref{eq:amplifierSDEq} (solved backward in time via the Euler–Maruyama method), the \emph{forward guidance} model in Eq.~\eqref{eq:amplifier_micro} (solved forward in time via Euler–Maruyama with guidance), and the \emph{mean-field} model in Eq.~\eqref{eq:NsystemSDE} for $N=100$ and $N=10,000$ (solved forward in time via $N$-dimensional coupled Euler–Maruyama).
    For all cases, 10 trajectories are shown.
    As is evident from the visualization, these formulations produce the same stochastic trajectories, with the mean-field model producing better agreement as $N$ increases.
    Simulation code is available at \url{https://github.com/dewitt-lab/Q-functionology}
    .}
    \label{fig:simulations}
\end{figure*}

We now illustrate the preceding comparison between reverse-time and forward-time diffusion hidden variable models in a concrete example (see also the numerical results of Figure~\ref{fig:simulations}).
Drummond and Reid \citep{DrRe20} consider the ``amplifier" dynamics as a toy model of measurement given by the Hamiltonian
\begin{align}
\label{eq:amplifier_dynamics}
H = -\frac{i\hbar^2}{2}\left(x\partial_x + \frac{1}{2}I\right),
\end{align}
which they argue leads to the Fokker-Planck type equation for the Husimi function:
\begin{equation}
\label{eq:amplifier_FP}
    \partial_t \Q = p\partial_p\Q - q\partial_q \Q + \partial_p^2\Q - \partial_q^2\Q.
\end{equation}
In other words, the Hamiltonian in Eq.\ \eqref{eq:amplifier_dynamics} gives rise to drift coefficients $a_1 = -p$, $a_2 = q$ and constant diffusion coefficient $b=\sqrt{2}$.
The ``retrocausal" diffusion microdynamics in Drummond and Reid's model is
\begin{align}
    \label{eq:amplifierSDEp}p(t) &= p(t_0) - \int_{t_0}^tp(t')\odif{t'} + \sqrt{2}\int_{t_0}^t \odif{w_1}\\
    \label{eq:amplifierSDEq}q(t) &= q(t_f) - \int_{t}^{t_f} q(t')\odif{t'} + \sqrt{2}\int_t^{t_f}\odif{w_2}.
\end{align}
As in the general case, one of the microdynamical equations describes evolution determined by a future boundary condition.

The alternative forward-time stochastic guidance model proposed in this letter keeps Eq.\ \eqref{eq:amplifierSDEp}, but reverses the time in Eq.\ \eqref{eq:amplifierSDEq} to obtain
\begin{equation}
\label{eq:amplifier_micro}
    q(t) = q(t_0) + \int_{t_0}^t \left(q(t') + 2 \frac{\partial_q \mu_{t'}(q(t'))}{\mu_{t'}(q(t'))}\right)\odif{t'}
    + \sqrt{2} \int_{t_0}^t \odif{\overline{w}_2}.
\end{equation}
These dynamics now involve a drift provided by a guidance term.

Next, we will show that the forward-time guidance dynamics in Eq.\ \eqref{eq:amplifier_micro} arises as the mean-field limit of a weakly interacting particle system.
The Fokker-Planck Eq.\ \eqref{eq:amplifier_FP} has a solution of the form
\begin{align}
\label{eq:amplifierFPsoln}
    \Q(p,q) = \frac{1}{2\pi}\exp\left(-\frac{1}{2}\left(\left(q-q_0e^{t}\right)^2 + \left(p-p_0e^{-t}\right)^2\right)\right)
\end{align}
for some initial value $(p_0,q_0)$.
Eq.\ \eqref{eq:amplifierFPsoln} describes a product of Gaussians with fixed variance and independently evolving means, each mean obeying classical dynamics.
In this case, we have the marginal distribution
\begin{align}
    \mu_t(q) = \frac{1}{\sqrt{2\pi}}\exp\left(-\frac{1}{2}\left(q-q_0e^{t}\right)^2\right),
\end{align}
which gives rise to the guidance term 
\begin{align}
\label{eq:amplifier_guidance}
\frac{\partial_q \, \mu_t\left(q\right)}{\mu_t(q)} = \partial_q\log\mu_t(q) = -\left(q-q_0e^{t}\right).
\end{align}
We provide a visualization in Fig.~\ref{fig:simulations} of stochastic trajectories for the amplifier dynamics with initial distribution given by Eq.~\eqref{eq:amplifierFPsoln}, allowing comparison of the ``retrocausal" model with the forward guidance model and $N$-system approximation.

We now seek to better understand the guidance drift term as a particular mean field interaction in an $N$-system.  To do so, we treat the additional drift term as arising from a force, written in terms of a potential.

We claim that one may identify the guidance term as arising from the pairwise harmonic potential
\begin{align}
    \label{eq:amplifier_potential}
    V(q,x) = \frac{1}{2}\left(q-x\right)^2.
\end{align}
Notice that this given pairwise interaction reproduces the forward-time evolution with the drift term in Eq.~\eqref{eq:amplifier_guidance}.
It follows from Eq.~\eqref{eq:amplifier_potential} that $\partial_q\log\mu_t$ is equal to the expectation value of $-\partial_qV$ relative to $\mu_t$:
\begin{align}
    \langle -\partial_q V(q,\cdot) \rangle_{\mu_t}\nonumber
    &= \int -\partial_q V(q,x)d\mu_t(x)\nonumber\\&= -\left(q-\langle q \rangle_{\mu_t}\right) \nonumber\\
&= \partial_q \log\mu_t(q).
\end{align}
The form of the harmonic potential makes apparent that the guidance term generally gives rise to a force on the particle toward the mean of the distribution, counteracting the diffusive behavior.

In the mean-field interpretation, we can understand the restorative force as arising in the $N\to\infty$ limit from a pairwise potential for an $N$-system.  To see this, consider again the guidance drift term in the $N$-system specified in Eq.~\eqref{eq:interactingSDE}, which in the amplifier system has the form $\partial_q\log\tilde{\mu}_t^N$.
To reproduce this drift term in the $N$-system, we define the mollified pairwise interaction potential
\begin{equation}
    \tilde{V}^N(q,x) = \left(V(q,\cdot)\ast K^N\right)(x) = \int V(q,y)K^N(x-y)\odif{y}.
\end{equation}
Then, since for each fixed $q$ and $y$, $V(q,x)K(x-y)$ vanishes fast enough as $x\to \infty$, one has
\begin{align}
    \int V(q,x) \tilde\mu^N_t(x) \odif{x} &= \int\int V(q,x) K^N(x-y) \mu_t^N(y)\odif{y}\odif{x}
    \nonumber\\&= \int \tilde{V}^N(q,y)\mu_t^N(y)\odif{y},
\end{align}
which implies
\begin{equation}\langle V(q,\cdot) \rangle_{\tilde\mu^N_t} = \langle \tilde V^N(q,\cdot) \rangle_{\mu^N_t}.
\end{equation}
It follows that $\tilde V^N \to V$, and hence
\begin{align}
    |- \partial_q\tilde{V}^N-\partial_q\log\tilde\mu^N_t |\to 0,
\end{align}
with convergence in distribution as $N\to\infty$.
Thus, the guidance term for the focal particle in Eq.\ \eqref{eq:interactingSDE} is approximated for large $N$ by the force from an average energy of pairwise interactions with all particles of the $N$-system.  In other words, one can interpret the restorative mean-reverting force in the bulk harmonic potential as arising from attractive pairwise interactions among particles in the mean-field limit of the $N$-system.

Note that the mean-reverting form of the guidance term can recover a mean-field interpretation for the dynamics of a wide range of Hamiltonians.  The guidance term will arise from a pairwise harmonic potential whenever the Fokker-Planck equation for $\Q$ admits a solution that is a Gaussian of fixed variance, but whose mean evolves according to a classical dynamics---for example, Gaussian solutions of $\Q$ to the free particle Fokker-Planck Eq.\  \eqref{eq:free_FP}.  We conjecture this holds more generally for Hamiltonians up to quadratic in the position and momentum operators whose Fokker-Planck equation contains no coupling between $p$ and $q$ in the drift terms.

\section{Conclusion}

We have shown that the ``retrocausal" hidden variable model of Drummond and Reid \citep{Dr21,DrRe20,DrRe21} is equivalent to a set of forward-time SDEs.
In the forward-time formulation, results from Anderson \citep{An82} allow one to recover the reverse-time diffusion by adding an extra force, expressed as a guidance term determined by a distribution that solves the Fokker-Planck equation.
We showed that the guidance term can be understood as a restorative force \emph{toward the mean} of the distribution.
The qualitative picture is that while the reverse-time diffusion model involves dissipation away from the mean, the forward-time equivalent compensates with precisely the force required to move the particle back along the same path.

We showed that this qualitative picture is born out exactly in the mean field limit, so that one can think of the particle as interacting with an ensemble via a restorative coupling to the (smoothed) empirical measure.
In this sense, the phase-space pilot wave of our model loses it's mystery; it can be understood as only a convenient summary arising in the $N\to\infty$ limit of averaging over pairwise interactions of particles in an ensemble.
Our results have practical implications: both the forward-time equations and the $N$-system approximation give rise to new methods for computing the trajectories allowed by Drummond and Reid's model, even for systems where solving the reverse-time equations is a challenge.

Likewise, our results have significant interpretive implications.
We have shown that models with a reverse-time SDE give rise to an equivalent forward-time guidance SDE---a stochastic pilot wave model generalizing the well-known de Broglie-Bohm model to cases where the guiding distribution is \emph{not} the wavefunction, but rather some phase space distribution determined by the wavefunction.  Further, we have shown that such a guidance SDE (with drift depending on a probability distribution) can be approximated arbitrarily well by a system of $N$ interacting particles (for large enough $N$) satisfying weakly coupled SDEs through the empirical measure---a model resembling and generalizing a type of many-worlds interpretation.  Together, these results express a tight relationship between Drummond and Reid's model and two broad, popular classes of interpretations of quantum mechanics: pilot wave interpretations and many-worlds interpretations.
Thus, for stochastic hidden variable models, we find a general trade-off between ``retrocausality" (captured by reverse-time equations), the guiding wave picture, and the mean-field interaction picture.

\section*{Acknowledgments}
The authors thank Chris Burdzy, Simon Friederich, Colin LaMont, Mritunjay Tyagi, and the participants in the UW Probability Seminar for helpful discussion and feedback on this work.  We also thank two anonymous reviewers for detailed comments that improved this letter.

\bibliography{bibliography.bib}

\end{document}